\documentclass[fleqn,usenatbib]{mnras}

\usepackage{newtxtext,newtxmath}
\usepackage[T1]{fontenc}
\usepackage{ae,aecompl}

\usepackage{graphicx}	

\title[Growth of structure VIPERS to SDSS]{Measuring the growth of structure by matching dark matter haloes to galaxies with VIPERS and SDSS}

\author[B. R. Granett et al.]{Benjamin R. Granett,$^{1}$\thanks{benjamin.granett@unimi.it}
Ginevra Favole,$^{2}$\thanks{gfavole@sciops.esa.int}
Antonio D. Montero-Dorta,$^{3}$\thanks{amonterodorta@gmail.com}
\newauthor
Enzo Branchini,$^{4,5,6}$  Luigi Guzzo,$^{1,7,8}$ Sylvain de la Torre$^{9}$
\\
$^{1}$Universit\`{a} degli Studi di Milano, Via Celoria 16, Milan 20133, Italy\\
$^{2}$European Space Astronomy Centre (ESAC), 28692 Villanueva de la Ca\~nada, Madrid, Spain\\
$^{3}$Departamento de F\'isica Matem\'atica, Instituto de F\'isica, Universidade de S\~ao Paulo, Rua do Mat\~ao 1371, CEP 05508-090, \\
S\~ao Paulo, Brazil \\
$^{4}$Department of Mathematics and Physics, Roma Tre University, Via della Vasca Navale 84, I-00146 Rome, Italy \\
$^{5}$INFN -- Sezione di Roma Tre, via della Vasca Navale 84, I-00146 Rome, Italy \\
$^{6}$INAF -- Osservatorio Astronomico di Roma, via Frascati 33, I-00040 Monte Porzio Catone (RM), Italy\\
$^{7}$INAF -- Osservatorio  Astronomico  di  Brera,  Via  Brera  28,  20122 Milan, via E. Bianchi 46, I-20121 Merate, Italy\\
$^{8}$INFN -- Sezione di Milano, Via Celoria 16, 20133, Milan, Italy\\
$^{9}$Aix Marseille Univ, CNRS, CNES, LAM, Marseille, France \\
}

\date{Accepted XXX. Received YYY; in original form ZZZ}

\pubyear{October 2019}
\volume{489}
\pagerange{653--662}

\begin{document}
\label{firstpage}
\maketitle

\begin{abstract}
We  test  the  history  of  structure  formation  from  redshift  1  to  today
by  matching  galaxies from  the  VIMOS  Public  Extragalactic  Redshift
Survey  (VIPERS)  and  Sloan  Digital  Sky Survey (SDSS) with dark matter
haloes in the MultiDark, Small MultiDark Planck (SMDPL), N-body simulation. We
first show that the standard subhalo abundance matching (SHAM) recipe
implemented with MultiDark fits the clustering of galaxies well both at
redshift 0 for SDSS and at redshift 1 for VIPERS. This is an important
validation of the SHAM model at high redshift. We then remap the simulation
time steps to test alternative growth histories and infer the growth index
$\gamma=0.6\pm0.3$. This analysis demonstrates the power of using N-body
simulations to forward model galaxy surveys for cosmological inference. The
data products  and  code  necessary  to  reproduce  the  results  of  this
analysis  are  available  online
(\url{https://github.com/darklight-cosmology/vipers-sham}).
\end{abstract}




\section{Introduction}

The growth of structure over cosmic time is a fundamental observable that
informs us about the expansion history and the physics of gravitational
instability, both of which are key ingredients for interpreting cosmic
acceleration \citep[e.g.][]{Huterer2015}. Surveys that map the distribution of
galaxies out to high redshift provide important measurements of the statistics
of the matter field and its evolution.   In the standard paradigm galaxies
form inside massive dark matter clumps, and these clumps build up
hierarchically \citep{White1991}.  The formation of dark matter structures and
their spatial statistics have been well investigated analytically and in
$N$-body simulations \citep[e.g.][]{BBKS1986,Springel2005}.

However, the connection between the galaxies detected in surveys and the
underlying matter distribution is complex \citep{Baugh2013,Wechsler2018}.
Observations show that the two-point clustering statistics  depend strongly on
the luminosity, colour, morphology and other physical properties of the galaxy
sample \citep{Davis1976,Giovanelli1986,Guzzo1997,Norberg2002,Zehavi2005,Polo2006,Marulli2013,Cappi2015,DiPorto2016} since
these properties are tied to the density environments the galaxies are found
in \citep{Blanton2007,Davidzon2016,Cucciati2017}.  These dependencies are
encoded in the galaxy bias $b_g$ that relates the two-point clustering
statistics of the galaxies to that of the underlying matter on large scales: $\xi_g(r,z) =
b_g(z)^2 \xi(r, z)$ \citep{Kaiser1984}.   It is the usual practice to
parametrize the bias function and marginalise over these parameters in a
cosmological analysis since they depend on the galaxy sample and the
peculiarities of the survey selection function \citep{Rota2017,Alam2017}.
Other approaches have been developed to infer the biasing function
using statistics of the galaxy distribution. \citet{DiPorto2016} constrain the
bias by matching the galaxy density distribution measured in a galaxy survey
with the distribution of dark matter in an $N$-body simulation assuming a
one-to-one correspondence.  We will follow a similar approach in this analysis
using dark matter haloes.

The process of matching the dark matter haloes in a simulation to the distribution of
galaxies selected by luminosity or stellar mass in a survey known as sub-halo
abundance matching (SHAM)  provides a simple yet
accurate prediction of galaxy bias
\citep{Vale2004,Conroy2006,Behroozi2010,Moster2010,Trujillo2011}.
The method requires an $N$-body simulation with sufficient resolution to identify and follow
the substructure within dark matter haloes \citep{Guo2014}.  \citet{Reddick2013} demonstrate
that a single halo property is sufficient to assign galaxies and that the
implicit choice of this property primarily affects the clustering on small
scales below 1$h^{-1}{\rm  Mpc}$.
Stochasticity or scatter in the relationship between the
 halo mass  and the galaxy luminosity has been shown to be less important when the galaxy sample
is sufficiently deep such that it is complete down to the characteristic
flattening of the luminosity function (or stellar mass function) \citep{Conroy2006,Reddick2013}.

At low redshift, spectroscopic surveys including the Two-degree Field Galaxy Redshift
Survey (2dFGRS), the Sloan Digital Sky Survey (SDSS) Main galaxy sample and the
Galaxy And Mass Assembly (GAMA) survey have appropriately broad and deep selection
functions.  At higher redshift, the VIMOS Public Extragalactic Redshift Survey
\citep[VIPERS,][]{Guzzo2014,Scodeggio2018} is unique with a cosmologically
representative volume.

The accuracy of SHAM to model galaxy clustering over cosmic time was first
demonstrated by \citet{Conroy2006} who compiled galaxy clustering measurements
to $z\sim5$.  \citet{Conroy2006}
developed a SHAM model to assign galaxy luminosities to haloes using the
equivalent of the halo property $V_{peak}$ that we define below. No additional
free parameters such as stochasticity or scatter in the assignment were used.
The success of \citet{Conroy2006} has motivated the development of the SHAM
model that we adopt  to describe the clustering of galaxies in SDSS and VIPERS.

The application of SHAM without free parameters is attractive for making
cosmological predictions.  For example,  \citet{He2018} extended SHAM to
modified gravity models and tested the validity of these models against the
standard $\Lambda$ cold dark matter scenario using galaxy clustering statistics.
To extend this technique more generally to constrain cosmological
parameters requires a large number of simulations that span a range of
cosmological models \citep{Harker2007}.   However, a practical shortcut can be taken to
avoid this computational expense.  It has been shown that a
simulation run in one model can be made to quantitatively look like a
simulation run in a different model by re-scaling the time and spatial
dimensions to match the expansion and growth histories
\citep{Angulo2010,Mead2014a,Mead2014,Mead2015,Zennaro2019}.  This approach was implemented
in a cosmological analysis pipeline by \citet{Simha2013}.

We apply the re-scaling algorithm here in a simplified context in
which we vary only the growth history quantified by $\sigma_8$, the variance
of the linear matter field on 8$h^{-1}{\rm Mpc}$ scales.    In practice,
modifying the evolution of $\sigma_8(z)$ in a simulation requires only
re-labeling the redshift of the outputs.  Using the MultiDark $N$-body
simulation \citep{Klypin2016}, we employ a parameter-free SHAM model to
predict the galaxy correlation function and directly constrain $\sigma_8(z)$
using measurements at redshift $z<0.106$ in SDSS and at redshift $0.5<z<1$ in
VIPERS. \cite{Harker2007} made a similar analysis on SDSS that employed
semi-analytic models for galaxy formation to predict the amplitude of galaxy
clustering.  \citet{Simha2013} carried out a full cosmological analysis using
SDSS making use of re-scaled simulations and SHAM.  We present
the preliminary application of these techniques to higher redshift.

The growth history $\sigma_8(z)$ may be parametrized by the growth index
$\gamma$ as \citep{Wang1998}
\begin{equation}\label{eq:growth}
\sigma_8(z) = \sigma_8(0) \exp\left[-\int_0^{z} \Omega_m(z')^{\gamma} d\ln (1+z')\right].
\end{equation}
The growth index in the standard model is $\gamma=0.55$.  Other parametrizations have been proposed more recently \citep[e.g.][]{Silvestri2013}; however, the use of the growth index
neatly separates the dependence on the expansion history given by $\Omega_m(z)$
from modifications to the gravity model \citep{Linder2005,Guzzo2008,Moresco2017}.

In this paper we first present a validation of the SHAM model over the
redshift range $0<z<1$ using well-characterised galaxy samples from SDSS and
VIPERS (Sec. 2, 3).  To give an additional test of the underlying assumptions
we select galaxies by luminosity and stellar mass with matching number
densities so that they share the same SHAM prediction.  We study systematic
errors arising from incompleteness and scatter in Sec. 4.  After demonstrating
the robustness of the SHAM model, we apply the re-scaling algorithm to the
MultiDark simulation and infer the cosmological growth of structure (Sec. 5).
Section 6 concludes with a discussion of the results.

\section{Galaxy redshift surveys}
\begin{figure*}
    \centering
    \includegraphics[width=2.3in]{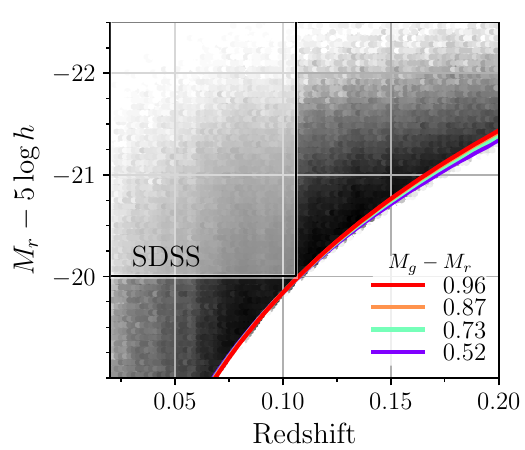}\includegraphics[width=2.3in]{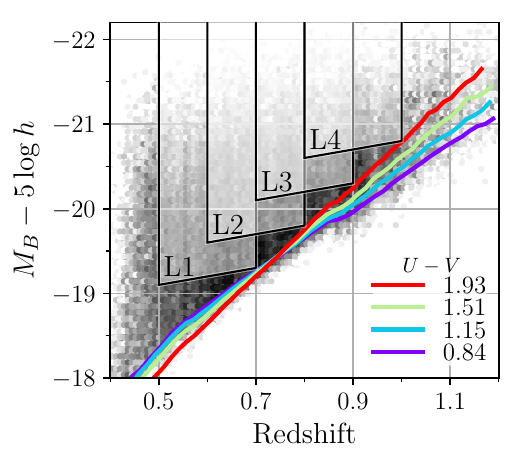}\includegraphics[width=2.3in]{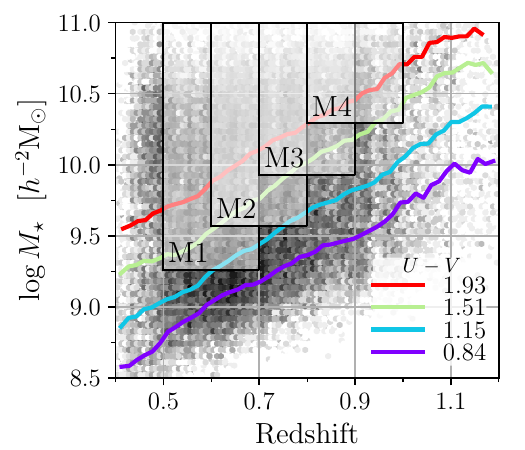}
    \caption{The SDSS and VIPERS samples used in this study.  Left: The selection of the SDSS sample on the absolute magnitude in the $r$ band, $M_r$.  Middle: the selection of the VIPERS luminosity samples on $M_B$ with an evolution trend.  Right: the selection of the VIPERS stellar mass samples.  In each panel the 90\% completeness limits are indicated by the lines as a function of the galaxy colour from blue to red (the colour is $M_g-M_r$ for SDSS and $U-V$ for VIPERS).}
    \label{fig:samples}
\end{figure*}
\begin{figure*}
\centering
    \includegraphics[width=6.5in]{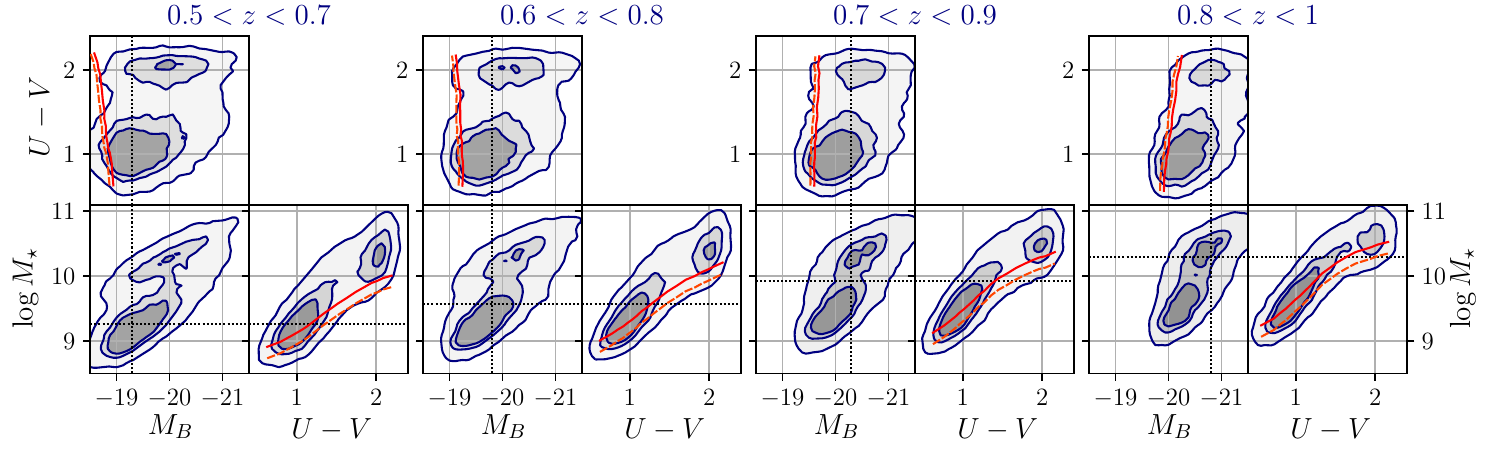}
    \caption{The distribution of the VIPERS sample as a function of $U-V$ colour, absolute magnitude $M_B$ and stellar mass for the four redshift bins. The contours contain 25, 50 and 90\% of the sample.  The horizontal and vertical dotted lines mark the stellar mass and absolute magnitude thresholds respectively of the subsamples used in the analysis.  The solid and dashed curves indicate the 90 and 50\% completeness limits in stellar mass and absolute magnitude as a function of colour.  The red sequence is above the stellar mass completeness limit in each redshift bin.}
    \label{fig:comp}
\end{figure*}
\begin{table*}
\centering
\begin{tabular}{c|c|r|l|r|r|r}
\hline
Sample     & Redshift       & Mean z & Threshold  & Count &  Volume  & Density\\
          &                &           &  &       & $10^6 h^{-3}{\rm Mpc}^3$ & $10^{-3} h^{3}{\rm Mpc}^{-3}$ \\
\hline
\hline
SDSS       & $0.020 < z < 0.106 $ & 0.063  & $M_r < -20.0$  &   117959    &  21.90     &   5.85   \\
\hline
\hline
L1         & $0.5 < z < 0.7 $ & 0.61 & $M_B<-19.3 + (0.7-z)$ & 23352 & 4.93 & 11.8\\
M1         & $0.5 < z < 0.7 $ & 0.61 & $\log M_{\star}>9.26 h^{-2}{\rm M_{\odot}}$ & 22508 & 4.93 & 11.8\\
\hline
L2         & $0.6 < z < 0.8 $ & 0.70 & $M_B<-19.8 + (0.8-z)$ & 20579 & 5.98 & 8.57\\
M2         & $0.6 < z < 0.8 $ & 0.70 &  $\log M_{\star}>9.57 h^{-2}{\rm M_{\odot}}$ & 19577 & 5.98 & 8.57\\
\hline
L3         & $0.7 < z < 0.9 $ & 0.80 & $M_B<-20.3 + (0.9-z)$ & 13046 & 6.96 & 4.79\\
M3         & $0.7 < z < 0.9 $ & 0.80 &  $\log M_{\star}>9.93 h^{-2}{\rm M_{\odot}}$ & 12270 & 6.96 & 4.79\\
\hline
L4         & $0.8 < z < 1.0 $ & 0.90 & $M_B<-20.8 + (1.0-z)$ &  6305 & 7.86 & 2.13\\
M4         & $0.8 < z < 1.0 $ & 0.89 &  $\log M_{\star}>10.29 h^{-2}{\rm M_{\odot}}$ & 5881 & 7.86 & 2.13\\
\hline
\hline
\end{tabular}
\caption{The galaxy samples used in this study.  The number density is weighted to correct for
survey incompleteness.}
\label{tab:samples}
\end{table*}

\subsection{SDSS Main Galaxy Sample at z < 0.1}
The Sloan Digital Sky Survey (SDSS, \citealt{York2000}) main galaxy sample (MGS, \citealt{Strauss2002}) provides a flux-limited census of galaxies in the low-redshift universe. In this paper, we use the SDSS MGS Data Release 7 (DR7, \citealt{Abazajian2009}), which includes spectroscopy and photometry for 499,546 galaxies with Petrosian extinction-corrected r-band magnitude $r<17.77$ at $z<0.22$, over 7300 square degrees.

We obtain the MGS data from the NYU Value Added Galaxy Catalogs (NYU-VAGC\footnote{https://cosmo.nyu.edu/blanton/vagc/}, \citealt{Blanton2005}), which provide K-corrections, absolute magnitudes, completeness weights and the survey mask. We use the DR7 LSS catalog, which employs a more restrictive r-band cut at $r<17.6$ in order to ensure a homogeneous selection across the SDSS footprint. The absolute magnitudes
in the {\it{ugriz}} bands included in the LSS catalog are
K-corrected to $z_0=0.1$ using {\it{kcorrect}} \citep{Blanton2003}. By blue-shifting the rest-frame to $z=0.1$, the effect of the correction is minimised.

The NYU-VAGC provides all the elements needed to measure the SDSS correlation function, including survey mask, randoms, and galaxy weights. Following the procedure described in \cite{Favole2017}, the NYU-VAGC randoms are corrected for the variation of completeness across the SDSS footprint. This correction is performed by down-sampling the random catalogue with equal surface density in a random fashion using the completeness as a probability function (see Section 3 in \citealt{Favole2017} for more details).

We apply two different galaxy weights to correct for angular incompleteness. The fiber collision weight, w$_{fc}$, accounts for the fact that fibers on the same tile cannot be placed closer than 55 arcsec. These weights correspond to the total number of neighbours within a 55-arcsec radius of each MGS galaxy for which redshift was not measured due to fiber collisions (i.e., w$_{fc} \ge$ 0). The second weight, w$_{c}$, accounts for the redshift measurement success rate in the mask sector where each galaxy lies, so that w$_{c} \le$ 1. The average completeness of the MGS is $\sim$80$\%$ (see \citealt{MonteroDorta2009}). In the computation of the correlation function, each galaxy is counted as $(1+w_{fc}) w_{c}$ and each random as $w_c$ since we previously diluted the random catalogue using the $w_c$ measurement completeness.

We select a single sample in the redshift range $0.02 < z < 0.106$ by imposing an r-band absolute-magnitude threshold $^{0.1}{\rm M}_r < -20.0$.  The uncertainty on the SDSS clustering measurement is estimated from the covariance matrix of 200 jackknife resamplings with constant galaxy number density \citep{2016MNRAS.462.2218F}.

\subsection{VIPERS at 0.5 < z < 1}

The VIMOS Public Extragalactic Redshift Survey
\citep[VIPERS][]{Guzzo2014,Scodeggio2018} provides high-fidelity maps of the
galaxy field at higher redshift.   The survey measured 90 000 galaxies with
moderate resolution spectroscopy using the Visible Multi-Object Spectrograph
(VIMOS) at VLT.  Targets were selected to a limiting magnitude of
$i_{AB}=22.5$ in 24 square degrees of the CFHTLS Wide imaging survey.   The
low redshift limit was imposed by a pre-selection based upon color which
effectively removed foreground galaxies while providing a robust flux-limited
selection at $z>0.5$.

The completeness of the VIPERS sample with respect to the parent flux-limited
sample is well-characterised in terms of the target sampling rate (TSR) and
spectroscopic redshift measurement success rate (SSR) \citep{Scodeggio2018}.
Additionally, close pairs of galaxies could not be targeted due to slit
placement constraints leading to a drop in the correlation function at very
small scales $<1 h^{-1}{\rm Mpc}$.  We correct for this effect by up-weighting
pairs according to their angular separation when computing the correlation
function \citep[see][]{Pezzotta2017}.

The VIPERS sample has photometric measurements from the UV to infrared which
have been used to infer the luminosity and stellar masses of the galaxies
\citep{Davidzon2013,Fritz2014,Davidzon2016,Moutard2016}.  The absolute
magnitudes are presented assuming a standard flat cosmological model with
$\Omega_m=0.3$ and $h=1$, but note that we compute the the number density of the samples in the MultiDark cosmology for the SHAM analysis.  The distribution of the rest-frame magnitude
$M_B$ is shown in Fig. \ref{fig:samples}.

For the analysis we select four samples in overlapping bins of redshift with
thresholds in $M_B$.  These samples are labelled L1, L2, L3 and L4 and listed
in Table \ref{tab:samples}.  We impose an evolving luminosity limit to account
for the luminosity trend for a passively evolving stellar population as
applied in previous VIPERS analyses \citep[e.g.][]{Marulli2013}. The selection
threshold in a redshift bin $z_0 < z < z_1$ is specified as: $M_{limit} =
M_{z_1} + (z_1 - z)$.  We also construct matching samples selected by stellar
mass that have the same number density.  These samples are labelled M1, M2, M3
and M4. The number density is computed as the weighted sum to correct for TSR
and SSR.  The completeness limits as a function of luminosity, stellar mass
and colour are shown in Fig. \ref{fig:comp}.

We make use of the VIPERS mock galaxy catalogues to estimate the covariance of
the correlation function measurements.  These catalogues were built from the
Big Multidark $N$-body simulation \citep{Klypin2016}.  Galaxies were simulated
using the halo occupation distribution (HOD) technique calibrated to reproduce
the number density and projected correlation function of VIPERS galaxies in
bins of luminosity and redshift  \citep{delaTorre2013,delaTorre2017}.  In
total, 153 independent realisations of the full VIPERS survey are available.

For each VIPERS sample we select a comparable sample from the mock catalogues by
setting a threshold in luminosity that gives the same number density.   We
confirm that the projected correlation function of these mock samples
approximately matches the amplitude of the VIPERS  measurements.

\section{Matching with dark matter haloes}

We use the MultiDark $N$-body numerical simulation \citep{Klypin2016} to model
the distribution and evolution of dark matter haloes. We choose the Small MultiDark Planck
(SMDPL) box, of side length $400h^{-1}\rm{Mpc}$, containing a total of
3840$^3$ particles. The simulation assumes a $\Lambda$CDM cosmology
\citep{planck2014}, with parameters $h = 0.677$, $\Omega_m = 0.307$,
$\Omega_{\Lambda} = 0.693$, $n_s = 0.96$ and $\sigma_8 = 0.823$. Dark-matter
haloes (including subhaloes) were identified using the ROCKSTAR code
\citep{rockstar2013}.

We make the connection between galaxies measured in VIPERS or SDSS and haloes
from the SMDPL snapshots with sub-halo abundance matching (SHAM,
\citealt{Vale2004}).  The link to the simulated haloes is made using the peak
maximum circular velocity of the particles in the halo over its formation
history ($V_{peak}$). The $V_{peak}$ property characterizes the halo mass
before disruption processes occur and it has been demonstrated that this is
important for modelling the distribution of satellite galaxies.  Velocity is
used instead of virial mass because it is more robustly defined in
simulations.  For further details we refer the reader to
\citealt{Conroy2006,Trujillo2011,Reddick2013,Campbell2018}.

We select galaxies based upon a stellar mass (or luminosity) threshold.  Then,
within a single simulation snapshot we select haloes by setting a threshold in
$V_{peak}$ that results in an equal number density.  These haloes become the
mock galaxies for the analysis.  In Sec. \ref{sec:systematics} we test the impact
of scatter or stochasticity in the relationship between the halo and galaxy
properties.  However, our main results are derived without scatter and in
this case the SHAM model is determined solely by the densities of the  samples
listed in Table \ref{tab:samples}.

\begin{figure}
    \centering
    \includegraphics[width=3.2in]{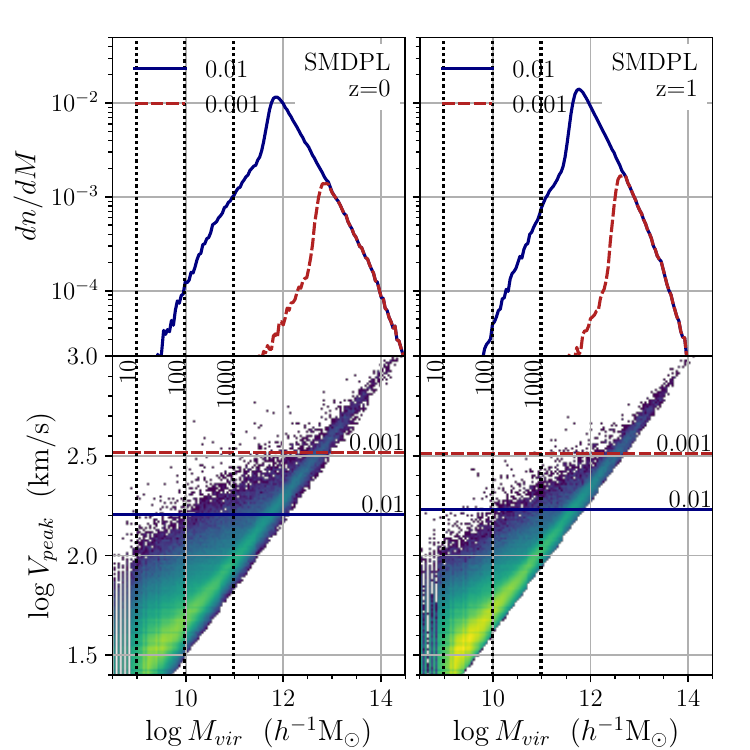}
    \caption{The distribution of $V_{peak}$ and $M_{vir}$ halo properties in SMDPL at $z=0$ (left panels) and $z=1$ (right panels).
    The horizontal solid and dashed lines in the bottom panels indicate thresholds in $V_{peak}$ that give number densities
    of $10^{-2}$ and $10^{-3} h^3{\rm Mpc}^{-3}$.  The $M_{vir}$ distributions after applying these selections are shown in the top panels (solid and dashed histograms).  The vertical dotted lines
    indicate the virial mass corresponding to 10, 100 and 1000 simulation particles.
    }
    \label{fig:halo}
\end{figure}

Fig. \ref{fig:halo} shows the distributions of haloes at $z=0$ and $z=1$ as a
function of virial mass $M_{vir}$ and $V_{peak}$.  Two $V_{peak}$ threshold
selections are indicated that give number densities $10^{-2}$ and $10^{-3}
h^3{\rm Mpc}^{-3}$. The median halo mass of the higher density selection is
$M_{vir}\sim7.5\times10^{11}h^{-1}{\rm M_{\odot}}$ at $z=1$ which corresponds
to 7500 simulation particles and guarantees that the haloes selected for the SHAM analysis are robustly defined \citep{Guo2014}.

\begin{figure}
    \centering
    \includegraphics[width=3.2in]{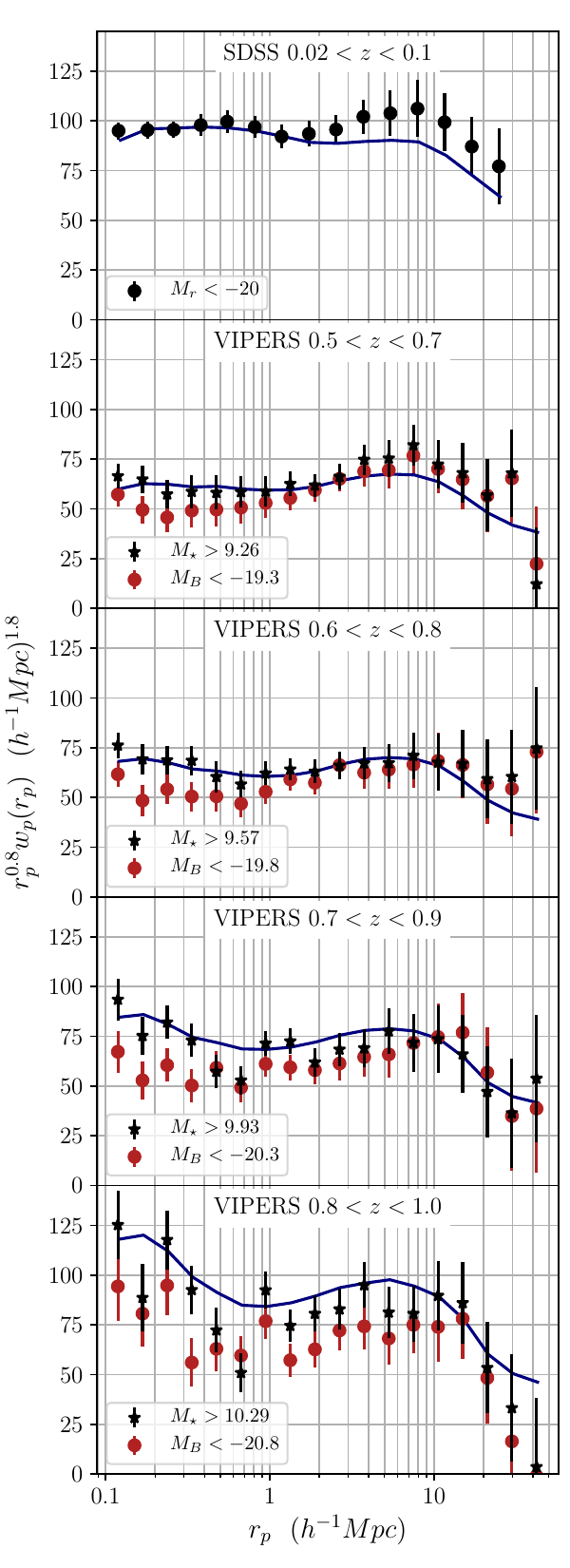}
    \caption{The projected correlation function measured in SDSS (top panel)
    and VIPERS (bottom four panels) in luminosity and stellar mass selected samples.
    The matched samples have the same number density and thus share the same SHAM model
    (solid curve).}
    \label{fig:cfbest}
\end{figure}

The clustering amplitude of the galaxy field can be inferred from measurements
of the projected correlation function without being strongly impacted by the
redshift-space distortion signal caused by peculiar velocities \citep{Davis1983}. The
projected correlation function $w_p$ depends on the perpendicular
separation $r_p$ and is computed by integrating along the line-of-sight ($\pi$
direction):
\begin{equation}
     w_p(r_p) = 2 \int_{0}^{\pi_{max}} \xi_g\left(r_p, \pi'\right) d\pi'.
\end{equation}
We set the integration limit to $\pi_{max}=50 h^{-1}{\rm Mpc}$.

We compute the redshift-space correlation function $\xi(r_p,\pi)$ for the
galaxy surveys using the Landy-Szalay estimator \citep{Landy1993}.  We employ
two correlation function code implementations, that of \citet{Favole2017} and CUTE
\citep{Alonso2012}.  The correlation functions of the MultiDark SHAM samples
are computed in the plane-parallel approximation taking advantage of the
periodic boundaries of the cubic simulation box.  The residual redshift-space
distortion signal in the projected correlation function is present both in the
galaxy and halo measurements, so we do not make any additional corrections.

We compute the projected correlation functions for the SHAM models at the
redshifts of the MultiDark snapshots, $w_p^{halo}(r_p,z|n)$, where $n$ is the
number density of the galaxy sample.  To compute the model between the
simulation snapshots at an arbitrary redshift we build a linear interpolation
function that is based on the principal component decomposition using the
first two eigenvectors.

Fig. \ref{fig:cfbest} shows the measured correlation function for each galaxy
sample and the corresponding SHAM model at the sample redshift.
There is good agreement between the SHAM model and the SDSS measurements.
 This confirms previous studies that developed and tested the
SHAM model on the SDSS galaxy correlation function \citep[e.g.][]{Reddick2013}.

 We find that the VIPERS luminosity selected samples have a clustering
amplitude that is  systematically lower  than the stellar mass selected
samples.  This discrepancy is more significant at smaller scales
$r_p<1h^{-1}Mpc$ and in the highest redshift bin.  In each redshift bin the
luminosity and stellar mass selected samples  were constructed to share the
same SHAM prediction, thus we find that the SHAM model better reproduces the
clustering of the stellar mass-selected sample.

The systematic difference in clustering amplitude between the luminosity
and stellar mass-selected samples is not unexpected since hydrodynamic
simulations have demonstrated that galaxy stellar mass is a better indicator
for the host halo mass \citep{ChavesMontero2016}.  We would expect the choice
to be less important when selecting galaxies based upon the rest-frame
luminosity in a redder band that is more tightly correlated to the stellar
mass \citep{Bell2001}.  This can explain the agreement with SHAM seen in SDSS
projected correlation functions for both $M_r$ selected and mass selected
samples\footnote{\citet{He2018} point out that the correlation functions of
luminosity and stellar mass selected samples are not similar in redshift space
and stellar mass should be preferred.} \citep{Reddick2013}.  On the other hand, the
bluer rest-frame band used in VIPERS (that is closest to the observed $i$ selection band) is more sensitive to recent star formation
activity and hence is less informative of the total mass of the galaxy.  The
consequence is that in VIPERS, the correlation function of galaxies selected
in $M_B$ is lower than for those selected by stellar mass.  This effect should
become more important at higher redshift as the rest frame for a fixed bandpass shifts to the blue and star formation activity becomes more
prevalent \citep{Haines2017}.

\section{Systematics}
\label{sec:systematics}

\begin{figure}
    \centering
    \includegraphics[width=3.2in]{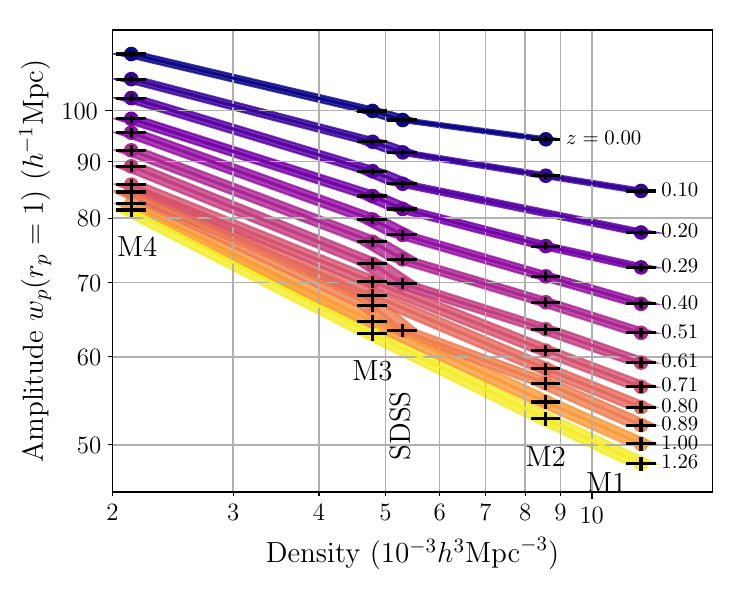}
    \caption{The correlation function amplitude at $r=1 h^{-1}{\rm Mpc}$ versus galaxy number density of all  SHAM samples used in our analysis. The error bars correspond to 5\% variation in number density, which is representative of the VIPERS sample variance.  In order to change the amplitude by 10\% requires a change of number density of 50\% at z=0 and 30\% at z=1.}
    \label{fig:density}
\end{figure}

\begin{figure}
    \centering
    \includegraphics[width=3in]{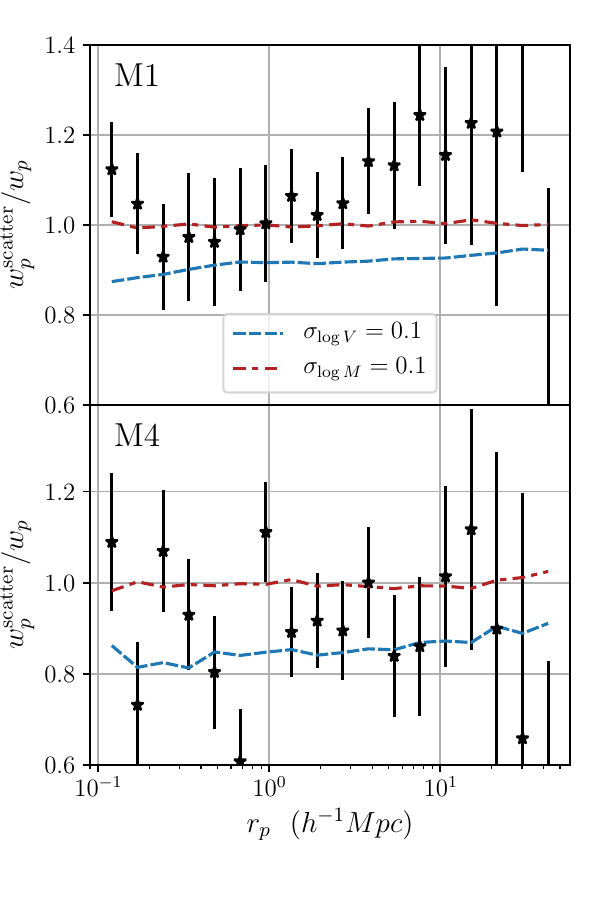}
    \caption{The relative change in the correlation function after introducing scatter in the SHAM procedure is shown for two mass-selected samples
    in VIPERS M1 $0.5<z<0.7$ (top) and M4 $0.8<z<1.0$ (bottom). A Gaussian scatter of 0.1 dex was applied to $M_{\star}$ (dash-dotted curve) or $V_{peak}$ (dashed curve). }
    \label{fig:scatter}
\end{figure}

We have found that the SHAM model can predict the galaxy clustering signal to
redshift 1; however, it is important to make note of the assumptions that have
been made and consider how extensions to the SHAM recipe would affect our
results.  On the observational side, uncertainty in the number density due to
sample variance or incompleteness propagates to the SHAM model as a systematic
error.  Fig. \ref{fig:density} summarises the SHAM models that we constructed
for this work and demonstrates the power-law relationship between the
clustering amplitude at $r=1 h^{-1}{\rm Mpc}$ and number density.
The horizontal error bars on this plot indicate 5\%  variations in number
density which is representative of the sample variance in the VIPERS samples.
The vertical error bar propagates this error to the amplitude of the
correlation function and is at the percent level.  In order to change the
amplitude of the correlation function by 10\% requires varying the number
density by 50\% at $z=0$ and 30\% at $z=1$.
These conclusions follow from the SHAM model that imposes that the
clustering amplitude is determined only by stellar mass (or other halo mass
proxy).  This is not precisely true since galaxy colour correlates with the
density of the environment at fixed stellar mass
\citep[e.g.][]{Davidzon2016}.

We also see from Fig. \ref{fig:density} that the SHAM prediction becomes less
sensitive to redshift at lower number density.  Therefore, to improve the
constraining power requires higher density samples which at high redshift
becomes observationally challenging.

The SHAM procedure can be extended to improve the precision of the
predictions. Scatter can be introduced to account for the fact that galaxies
of a specific stellar mass are associated with a greater variety of halo
properties than the SHAM dictates.  This may be due to stochastic processes or
error in the host halo assignment due to missing physical ingredients.
Investigations with hydrodynamic simulations indicate that the relationship
between galaxy stellar mass and halo $V_{peak}$ is approximately 0.1 dex
\citep{ChavesMontero2016}.

Fig. \ref{fig:scatter} shows the effect of scatter following two approaches.
First, we consider scatter applied to the stellar mass
(\citealt{Behroozi2010}, see also \citet{Trujillo2011} who apply scatter to luminosity).  From the observational perspective, this
scatter cannot be too large otherwise the intrinsic (deconvolved) stellar mass
function would be inconsistent with observations.  A large scatter also
requires extrapolating the stellar mass function to low masses below
observational limits.  We thus test scatter in stellar mass of 0.1 dex.   We
find that scatter of $\sigma_{\log M}=0.1$ dex has no effect on the measured
correlation function at the percent level for the number densities of the
VIPERS samples.  This is due to the fact that the stellar mass function is
flattening at the selection threshold \citep{Reddick2013}.

Next we consider a dispersion in $V_{peak}$.  This implies that $V_{peak}$ is
not a perfect proxy for galaxy assignment.  The advantage of applying scatter
to $V_{peak}$ is that a large scatter may be introduced without modifying the
stellar  mass function of galaxies.    We find that the scatter of
$\sigma_{\log V}=0.1$ dex does modify the amplitude of the correlation
function by 10 to 20\% in the VIPERS samples.  The scatter can improve the
match of the VIPERS data at high redshift but is not required given the
statistical error. However, scatter at the same level applied at lower
redshift is ruled out. The introduction of free parameters to account for
redshift-dependent scatter would greatly limit the cosmological
interpretation.

\section{Growth of structure}

\begin{figure}
    \centering
    \includegraphics[width=3.2in]{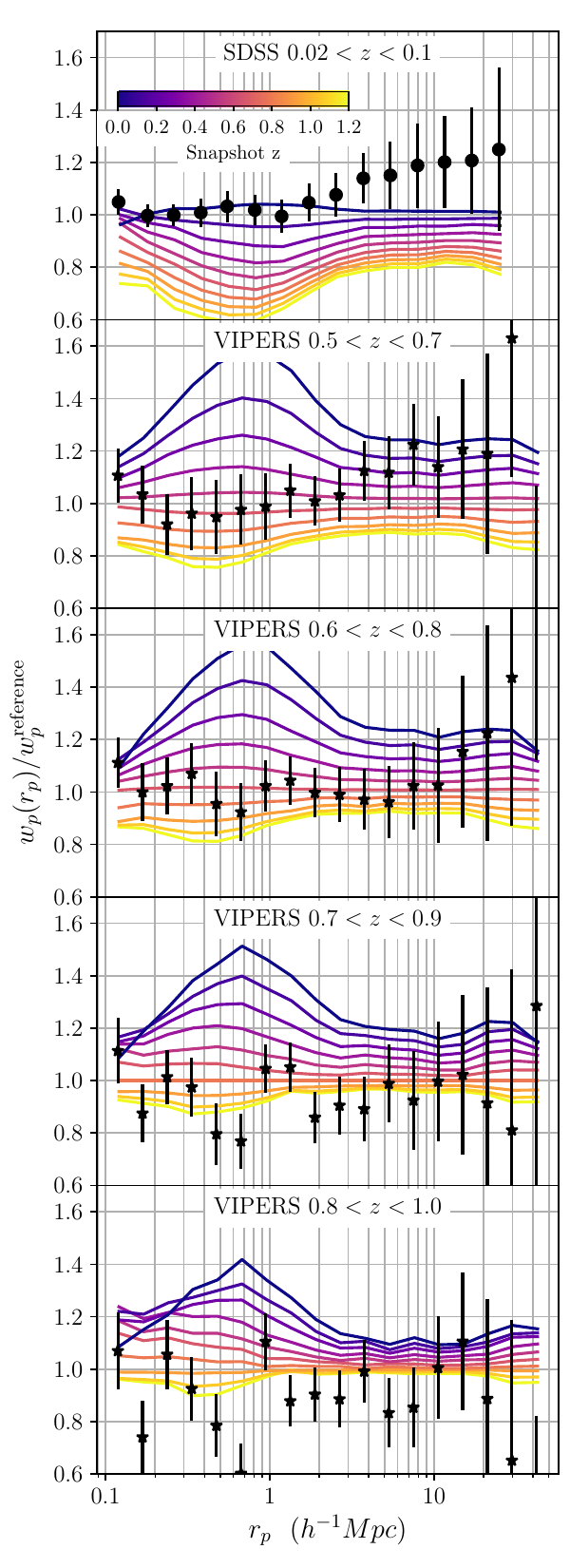}
    \caption{The SHAM model projected correlation functions computed over a range
    of simulation redshifts $0<z<1.2$.  In each panel the correlation function
    has been divided by the SHAM model at the sample redshift.  The data
    points indicate the SDSS sample (top panel) and VIPERS stellar mass
    selected samples (bottom four panels).}
    \label{fig:cf}
\end{figure}

\begin{figure*}
    \centering
    \includegraphics[width=6.5in]{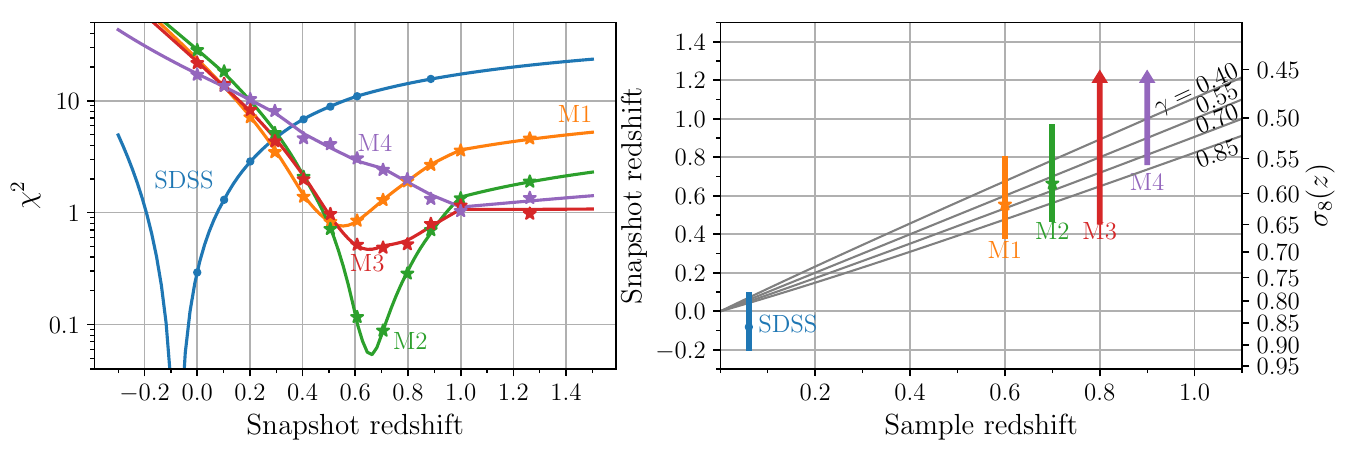}
    \caption{Left: the $\chi^2$ statistics for each galaxy sample (SDSS, M1, M2, M3, M4)
    as a function of redshift.  The markers indicate the SHAM models computed from the simulation snapshots while the curves were derived by linear interpolation of the models.  Negative snapshot redshifts (scale factor $>1$) corresponds
    to running the simulation into the future.
    Right: The best-fitting SHAM model as a function of its simulation snapshot is
    plotted for each galaxy sample shown on the left.  The right-most scale indicates the value of
    $\sigma_8(z)$ of the simulation snapshots.  Three alternative growth histories
    are over-plotted with growth index $\gamma=0.4, 0.7$ and 0.85 which give different mappings between
    the simulation redshift and the sample redshift.}
    \label{fig:fit}
\end{figure*}

\begin{figure}
    \centering
    \includegraphics[width=3in]{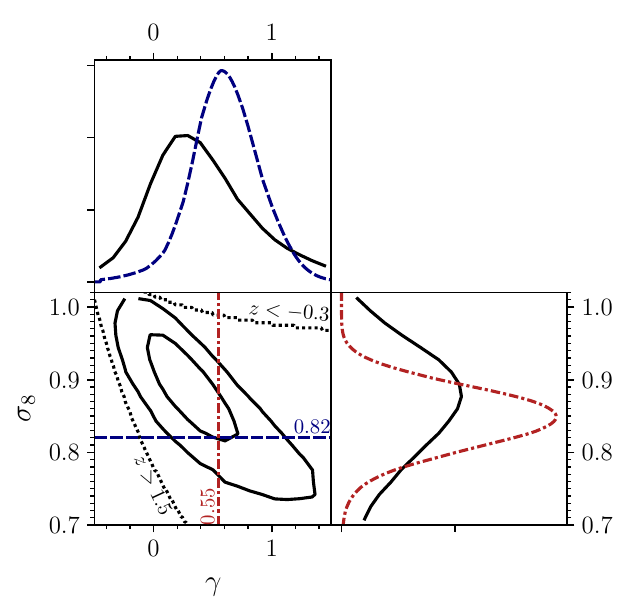}
    \caption{The likelihood degeneracy between the model parameters $\gamma$ and $\sigma_8$ today.  The contours mark the 1- and 2-$\sigma$
    levels.  The broken curves show the constraints on $\gamma$ with fixed $\sigma_8=0.82$ and
    on $\sigma_8$ with fixed $\gamma=0.55$.  The dotted curves indicate the borders of regions requiring extrapolation well beyond the simulation snapshots at $z<-0.3$ or $z>1.5$.}
    \label{fig:like}
\end{figure}

We now adopt the SHAM model without scatter to constrain the growth of structure.
 For each galaxy sample, we construct a halo sample with
matching number density for each one of 12 simulation outputs with snapshot
redshifts $0<z_{snap}<1.3$.  The correlation functions of the halo samples from each snapshot are over-plotted in the panels of Fig.
\ref{fig:cf}.

The best-fitting snapshot redshift was found for each sample by minimising the $\chi^2$ statistic
over redshift
\begin{equation}\label{eq:chi2}
\chi^2 = \sum_{i,j} \left(w_i^{obs} - w_i^{halo}(z)\right)C^{-1}_{ij}\left(w_j^{obs} - w_j^{halo}(z)\right)
\end{equation}
where $i$ and $j$ index the $r_p$ bins of the projected correlation function.
The analysis was made on scales greater than $r_{min}=1h^{-1}{\rm Mpc}$ to
avoid systematic uncertainties in both the observations and simulations.  The covariance matrices were inverted using the singular-value decomposition algorithm with a threshold of 0.1 on the relative size of the eigenvalues.
Fig. \ref{fig:fit} shows  the $\chi^2$ values and best-fitting redshifts.
The uncertainty of the determinations was estimated with the threshold
$\Delta \chi^2=1$.

The evolution of $\sigma_8$ is shown in Fig \ref{fig:fit} for alternative
gravity models parametrized by the growth index $\gamma$. The mapping is defined using
the growth equation $\sigma_8(z)$ (Eq. \ref{eq:growth}).
Considering the growth history in the MultiDark cosmology $\sigma_8^{MD}(z)$ and
an alternative model $\sigma'_8(z|\gamma)$ we determine the snapshot redshift $z_{MD}$
 that satisfies $\sigma_8^{MD}(z_{MD}) = \sigma'_8(z|\gamma)$.

In order to test models with high values of $\sigma_8$ we would need
simulation outputs at scale factors $a>1$ ($z<0$).  Since these are not
available in MultiDark, we linearly interpolate the correlation function to emulate
these outputs.  We also extrapolate to higher redshift which is required
to test models with low $\sigma_8(z)$.

We computed the joint likelihood defined by the $\chi^2$ in Eq. \ref{eq:chi2}
of each correlation function measurement as a function of $\sigma_8$ and
$\gamma$.  All other cosmological parameters were implicitly held fixed at the
fiducial values of the MultiDark simulation.  The likelihood surface is shown
in Fig. \ref{fig:like}.  Some regions of the parameter space require
extrapolation of the model well beyond the simulation snapshots. The limits
requiring extrapolation to $z<-0.3$ and $z>1.5$ are indicated by the dotted
curves in the figure but they are not excluded from the likelihood analysis.
The marginalised constraints are $\gamma=0.2^{+0.4}_{-0.3}$ and $\sigma_8=0.87\pm0.07$.
By fixing the value of $\sigma_8$ today to the MultiDark value $\sigma_8 = 0.82$ we find the growth index  $\gamma=0.6^{+0.3}_{-0.2}$. Considering the standard model
with $\gamma=0.55$ gives $\sigma_8=0.85\pm0.04$.

\section{Discussion and conclusions}

At low redshift the distribution of haloes has been shown to be a good proxy
for the distribution of galaxies and the SHAM recipe has been a success for
modelling galaxy clustering.  This is particularly true for galaxy samples that
are complete to the characteristic luminosity $L_{\star}$.   At higher
redshift VIPERS uniquely provides a dataset to complement low redshift
studies.   Here, we have found that the standard SHAM model without free
parameters reproduces the amplitude of the projected correlation function over
redshift range $0<z<1$ spanning SDSS and VIPERS.

We tested both luminosity and stellar mass selected sampled in VIPERS
constructed to have the same SHAM model.  The luminosity-selected samples were
found to have a lower clustering amplitude.  This supports the claim that
stellar mass is a better proxy for the host halo mass.  We expect that
luminosity becomes less informative at higher redshift due to the greater
influence of star formation activity particularly in bluer rest-frame
photometry.

Observational scatter in the relationship between stellar mass and the halo
$V_{peak}$ property cannot significantly impact the correlation function.  We
tested  scatter in stellar mass at the level of 0.1 dex and found no change in
the correlation function and greater levels of scatter is not consistent with
the observed shape of the stellar mass function. However, scatter
applied to $V_{peak}$ at the level of 0.1 dex does modify the amplitude of the
correlation function.

After demonstrating that SHAM can be successfully used to model the VIPERS
sample, we apply the re-scaling algorithm proposed by \citet{Angulo2010} to
test the history of structure formation.  The growth history provide direct
constraints on alternative cosmological models with modifications to gravity.
We estimate the growth index $\gamma$  to be $\gamma=0.6\pm0.3$ considering
SDSS and the VIPERS stellar mass selected samples.  The constraint was derived
 by fixing the value of $\sigma_8$ today.  Allowing $\sigma_8$ to vary
significantly reduces the constraining power of the data we consider.  The
sensitivity of the SHAM prediction depends on number density and we expect
that the precision measurements from upcoming photometric and spectroscopic
surveys at redshift $\sim1$ will allow robust constraints on both the
normalisation and the redshift dependence of $\sigma_8(z)$.

The constraints we find may be compared to those from previous studies based
on galaxy peculiar velocities and redshift-space distortions
\citep{Guzzo2008,Song2009,Blake2011,Beutler2012,Blake2013,Johnson2014,Alam2017}.
Galaxy velocities on large scales are sensitive to the derivative of the
growth factor $f = -d\log \sigma_8(z)/d\log(1+z) = \Omega_m(z)^\gamma$. Using
VIPERS data \citet{delaTorre2017} presented a 20\% measurement on $f\sigma_8$
in two redshift bins at $z=0.60$ and 0.86. Transforming these constraints to
$\gamma$ we find $\gamma=0.57^{+0.3}_{-0.4}$.  \citet{Hudson2012} compiled
measurements from peculiar velocity and redshift-space distortion surveys and
reported the joint constraint $\gamma=0.619\pm0.054$.  Using the BOSS and eBOSS
surveys \citet{Zhao2019} found $\gamma=0.469 \pm 0.148$.
By convention redshift-space distortion analyses fix the amplitude of
clustering at high redshift where it is constrained by measurements of the
cosmic microwave background.  In contrast our method is sensitive to the
integrated growth over a period at late time from redshift 1 to 0.

We investigated the effect of systematic uncertainties that would impact the
SHAM prediction through the dependence on number density.  The sample variance
present in the VIPERS sample propagates to the correlation function amplitude
at the percent level, and so cannot make a significant contribution to the
error.  Incompleteness at the ~30\% level would change the correlation
function amplitude by 10\%, but we have no evidence for the existence of such
a population of missing galaxies.  Fig. \ref{fig:comp} shows that the sample
is incomplete in stellar mass only for the reddest galaxies at high redshift.
A significant population of missing red galaxies could affect the
clustering amplitude and alter the trend with density shown in Fig.
\ref{fig:density}; however, we do not expect our results to be significantly
biased considering the level of precision of the VIPERS measurements at high
redshift.  Upcoming surveys such as ESA Euclid will target galaxies in the
near infrared and may shed additional light on the importance of stellar mass
incompleteness.

The SHAM recipe may be extended in future work to improve the precision of the
analysis. Scatter was not needed to fit the VIPERS data, but a degree of
intrinsic scatter is expected in the relationship between galaxy and halo
properties.  More flexible SHAM models can also be used to model samples that
suffer from incompleteness \citep{Favole2016,Rodriguez2017, Favole2017} or
completeness corrections can be inferred from deeper samples. Secondary
dependencies that are a signature of assembly bias such as the halo formation
time can also improve the precision of the SHAM model
(\citealt{Hearin2013,Miyatake2016,MonteroDorta2017,Niemiec2018,Lin2016}). The
additional parameters in these models may be degenerate with the cosmological
information we are attempting to extract, but there is a clear way forward if
they can be constrained from observations such as weak lensing measurements
\citep{Favole2016}.

\section*{Acknowledgements}
We thank Jianhua He for his expertise and helpful discussions and Gabriella De Lucia for making critical suggestions.
ADMD thanks FAPESP for financial support.
GF is supported by a European Space Agency (ESA) Research Fellowship at the European Space Astronomy Centre (ESAC), in Madrid, Spain.
EB is supported by MUIR PRIN 2015 ``Cosmology and Fundamental Physics: illuminating the Dark Universe with Euclid'', Agenzia Spaziale Italiana agreement ASI/INAF/I/023/12/0, ASI Grant No. 2016-24-H.0 and INFN project ``INDARK.''

We thank New Mexico State University (USA) and Instituto de Astrof\'isica de Andaluc\'ia CSIC (Spain) for hosting the Skies \& Universes site for cosmological simulation products.

This paper uses data from the VIMOS Public Extragalactic Redshift Survey (VIPERS). VIPERS has been performed using the ESO Very Large Telescope, under the ``Large Programme" 182.A-0886. The participating institutions and funding agencies are listed at \url{http://vipers.inaf.it}.



\bibliographystyle{mnras}
\bibliography{references}

\bsp	
\label{lastpage}
\end{document}